\documentclass[twocolumn,showpacs,preprintnumbers,amsmath,amssymb,prl]{revtex4-1}
\usepackage{stackengine,graphicx} 
\usepackage{epsfig} 
\usepackage{dcolumn} 
\usepackage{bm} 
\usepackage{xcolor}
\usepackage{xspace}
\usepackage{comment}
\usepackage{ulem}
\graphicspath{{./}{Figures//}}
\newcounter{saveenumi}

\newcommand{\be}{\begin{enumerate}}
\newcommand{\ee}{\end{enumerate}}
\newcommand{\rout}[1]{\textcolor{red}{\vphantom{#1}}}
\newcommand{\ins}[1]{\textcolor{black}{#1}}
\newcommand{\instwo}[1]{\textcolor{black}{#1}}

\definecolor{robgreen}{rgb}{0.333,0.62,0.18}
\definecolor{linebrown}{HTML}{bd6e00}
\definecolor{lineblue}{HTML}{0000ff}
\definecolor{linegreen}{HTML}{217200}
\definecolor{linered}{HTML}{d20d0d}

\def\cis{CuIr$_2$S$_4$}

\def\feses{FeSe}
\def\fese{FeSe}
\def\fd3m{Fd$\overline 3$m}
\def\p1bar{P$\overline 1$}
\def\i41amd{I4$_{1}$/amd}
\def\t2g{$t_{2g}$}

\newcommand{\pdfgui}{\textsc{PDFgui}\xspace}

\newcommand{\pdfgetxthree}{\textsc{PDFgetX3}\xspace}

\newcommand{\fittwod}{\textsc{Fit2D}\xspace}

\begin{document}
%
%
\begin{abstract}
We report pair distribution function measurements of the iron-based superconductor \fese\ above and below the structural transition temperature.
Structural analysis reveals a local orthorhombic distortion with a correlation length of about 4 nm at temperatures where an average tetragonal symmetry is observed.
The analysis further demonstrates that the local distortion is larger than the global distortion at temperatures where the average observed symmetry is orthorhombic.
Our results suggest that the low-temperature macroscopic nematic state in \fese\ forms from an imperfect ordering of orbital-degeneracy-lifted nematic fluctuations which persist up to at least 300~K.
\end{abstract}

\title{Room temperature local nematicity in FeSe superconductor}

\author{R.~J.~Koch$^{1,*}$, T.~Konstantinova$^{1,2}$, M.~Abeykoon$^{3}$, A.~Wang$^{1,\dag}$, C.~Petrovic$^{1}$, Y.~Zhu$^{1,2}$, E.~S.~Bozin$^{1}$, and S.~J.~L.~Billinge$^{1,4}$}

 \affiliation{$^{1}$Condensed Matter Physics and Materials Science Department, Brookhaven National Laboratory, Upton, NY~11973, USA}
 \affiliation{$^{2}$Department of Physics and Astronomy, Stony Brook University, Stony Brook, NY~11794, USA }
 \affiliation{$^{3}$Photon Science Division, Brookhaven National Laboratory, Upton, NY~11973, USA}
 \altaffiliation{rkoch@bnl.gov\\
 $^\dag$ Present address: School of Physics, Chongqing University, Chongqing 400044, China}
 \affiliation{$^{4}$Department of Applied Physics and Applied Mathematics, Columbia University, New York, NY~10027, USA}

\date{\today}
\maketitle

In high-temperature iron-based superconductors, nematicity, or $C_4$ to $C_2$ rotational symmetry breaking, is believed to be closely related to superconductivity as both are correlation-driven electronic instabilities which often coincide in the doping phase diagram~\cite{bohmer_nematicity_2018}.
The relationship is tantalizing but its nature remains unclear, as the two are frequently entwined with spin and orbital order~\cite{fernandes_what_2014, si_high-temperature_2016}.

The emergence of nematicity is marked by a tetragonal to orthorhombic global symmetry breaking structural transition at $T_s$~\cite{hsu_superconductivity_2008,hosoi_nematic_2016} as well as a lifting of the degeneracy of electronic orbitals in the material.
For example, in \fese , angle-resolved photoemission spectroscopy (ARPES) studies have demonstrated that the 3$d_{yz}$ and 3$d_{xz}$ Fe-Fe molecular orbitals take on a 50~meV energy difference below $T_o=90$~K.
The observation of orbital degeneracy lifting (ODL) coincides with orbital ordering (OO) which in turn accompanies or indeed even drives~\cite{baek_orbital-driven_2015} the global structural transition such that $T_o=T_s$~\cite{shimojima_lifting_2014, mcqueen_tetragonal--orthorhombic_2009}.

Although the OO and structural transition occur concurrently, a notable discrepancy exists between their energy scales.
The 50~meV orbital splitting corresponds to a temperature-scale of about 580~K rather than the 90~K of the transition. An alternative possibility is that the orbital degeneracy lifting happens at higher temperature, but the orbitals only order at $T_s$. This is similar to the case of \cis~\cite{bozin_local_2019}, where a short-range ordered (SRO) ODL state was discovered to exist well above the OO temperature.

The well established lattice symmetry of \fese\ above 90~K is tetragonal, with $C_4$ rotational symmetry about the c-axis.
The low-temperature orthorhombic lattice in this system has $C_2$ symmetry along the same axis~\cite{kothapalli_strong_2016}.
A structural signature for the existence of a SRO $C_2$ nematic phase above 90~K should manifest as a local, SRO orthorhombic (or similar) distortion, which does not order over long length scales.
The x-ray atomic pair distribution function (xPDF) measurement \ins{provides a histogram of interatomic distances in a material, and as such is well suited for probing such local distortions}\rout{is an ideal probe for such local distortions}~\cite{billi;cc04, egami;b;utbp12, bozin_local_2019}.
\rout{This has motivated the current work, where}\ins{Here} we use xPDF to study an \fese\ superconducting sample at temperatures both below and well above $T_o$ and $T_s$.

In analogy with observations of ODL imposed structural distortions in \cis~\cite{bozin_local_2019}, our xPDF study indicates a large orthorhombic distortion in \rout{the local structure of}\fese\ up to 300~K, well above $T_s$ where the average structure is tetragonal.
Surprisingly, even at 84~K, just below the 90~K transition temperature, the local orthorhombic distortion is significantly larger in magnitude than the average crystallographic orthorhombicity that is recovered over large length scales, suggesting an imperfect long-range ordering of the local ODL state below $T_s$.

The nature of the local symmetry breaking is consistent with local degeneracy lifting of 3$d_{yz}$ and 3$d_{xz}$ orbitals associated with Fe-Fe bond formation.
Our observation of a preformed local ODL broken symmetry state at high temperature portrays a picture where crystallization of the liquid-like \ins{orbital} state occurs at the global symmetry breaking structural transition at T$_s$.
This provides a rationale for the seemingly disparate energy scales involved.
It can also explain the orbital degeneracy breaking above T$_s$ implied by ARPES~\cite{zhang_observation_2015} and suggests that the ODL has a Jahn-Teller origin.


The \fese\ sample was synthesized through chemical vapor transport method using a eutectic mix of KCl and AlCl$_3$ as the transport agent~\cite{hu_synthesis_2011,chareev_single_2013}.
Sample quality was validated by a temperature dependent resistivity \rout{measurement}\ins{characterization} (Fig.~\ref{fig:transport} in Supplementary Information (SI)), which shows evidence for the structural phase transformation at $T_s=90$~K and a robust superconducting transition below $T_c=10$~K.
Synchrotron x-ray total scattering experiments were conducted at the 28-ID-1 (PDF) beamline at the National Synchrotron Light Source-II (NSLS-II) at Brookhaven National Laboratory (BNL).
The sample was loaded into a 1~mm inner-diameter kapton capillary and \rout{measured}\ins{data collected} at 300~K and 84~K using a liquid N$_2$ cryostream.
\rout{Data were collected}\ins{Measurements were carried out} in the rapid acquisition pair distribution function (RaPDF) mode~\cite{chupa;jac07}, with an x-ray energy of 74.69~keV ($\lambda = 0.1660$~\AA).
A two-dimensional (2D) PerkinElmer area detector was used, with a sample-to-detector distance of 204~mm determined by calibrating to a sample of known lattice parameter (Ni).

The 2D data were integrated and converted to intensity as a function of momentum transfer $Q$ using the software \fittwod~\cite{hamme;hpr96}.
The program \pdfgetxthree v2.0~\cite{juhas;jac13} was used to correct, normalize, and Fourier transform the diffraction data to obtain the experimental xPDF, $G(r)$ up to a momentum transfer of $Q_{max}=29$~\AA$^{-1}$ which was chosen as the best tradeoff between real-space resolution and noise in the data.

\begin{figure}[tbp]
\includegraphics[width=2.5in]{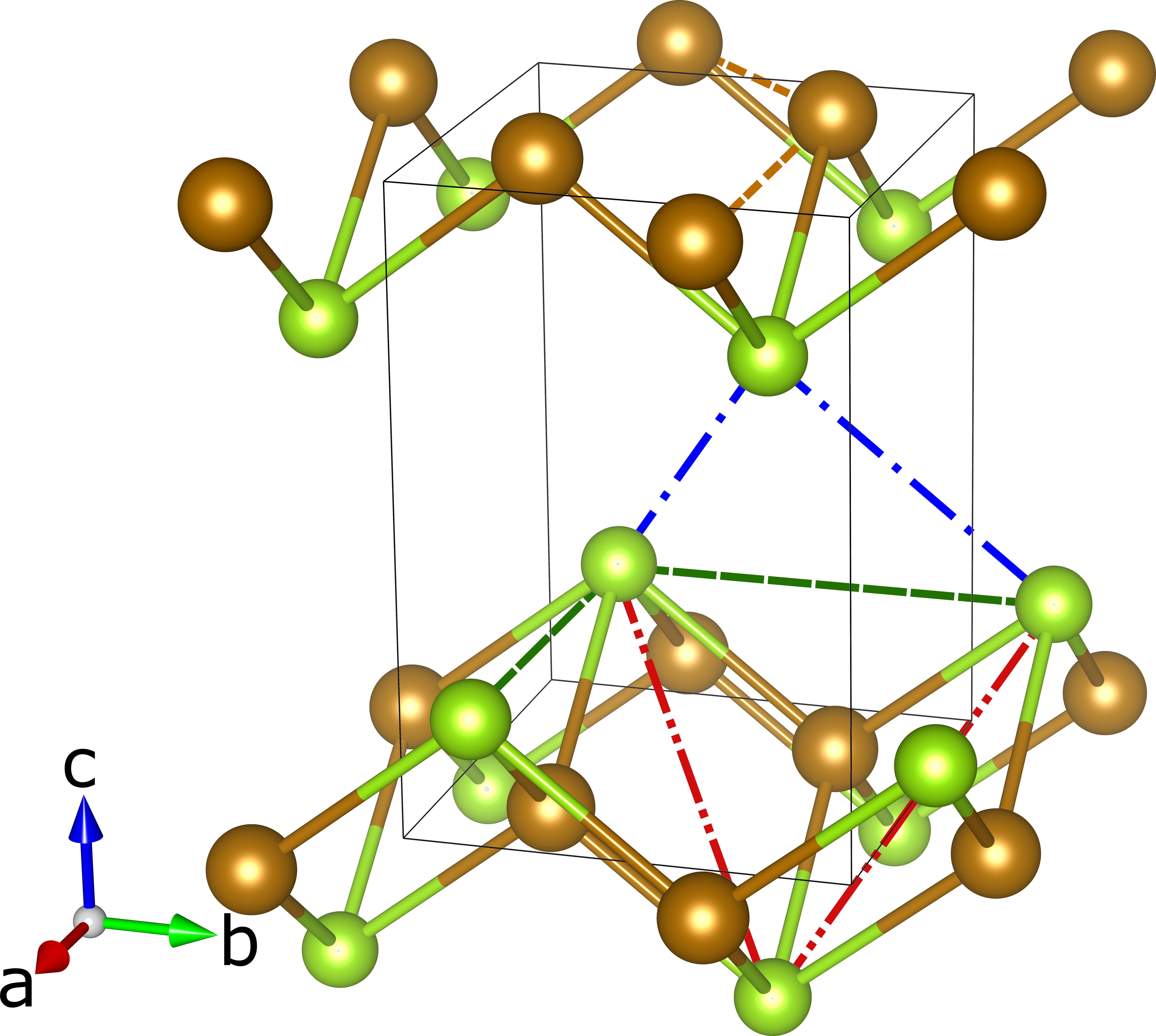}
\caption{\label{fig:strucfig} Perspective view of the FeSe structure with tetragonal symmetry.
Iron is shown in brown, and selenium in green.
Pair distances constrained by tetragonal symmetry to be identical are highlighted by dashed lines of the same color.
Tetragonal symmetry allows only a single Fe-Fe pair distance (brown dashed lines), and the orthorhomic distortion allows these distances to differentiate.
The three unique Se-Se pair distances shown by red, green, and blue dashed lines become five distinct Se-Se pairs within the orthorhombic symmetry.}
\end{figure}

The average structure for \fese\ is described by a tetragonal model at higher temperatures and an orthorhombic model at lower temperatures.
The tetragonal model ($P4/nmm$ space-group, shown in Fig.~\ref{fig:strucfig}) consists of FeSe slabs featuring a Fe square planar sublattice.
Each Fe is coordinated by four Se creating layers of edge-shared FeSe$_4$-tetrahedra, regularly stacked along the \textbf{c} lattice direction~\cite{hsu_superconductivity_2008,hosoi_nematic_2016}.
In this model, the asymmetric unit contains one \fese\ formula unit, with Fe at $(0.25, 0.75, 0.0)$ and Se at $(0.25, 0.25, z)$.
The lower temperature orthorhombic model ($Cmma$ space-group) is related to the tetragonal model through a rotation in the \textbf{a}-\textbf{b} plane and a rectangular distortion of the Fe sublattice~\cite{bohmer_nematicity_2018}.
Here, Fe sits at $(0.25, 0.0, 0.5)$ and Se at $(0.0, 0.25, z)$.

The xPDF data were fit with the \pdfgui~\cite{farro;jpcm07} program using these two models (see SI for details).

\begin{figure}[tbp]
\includegraphics[width=3.5in]{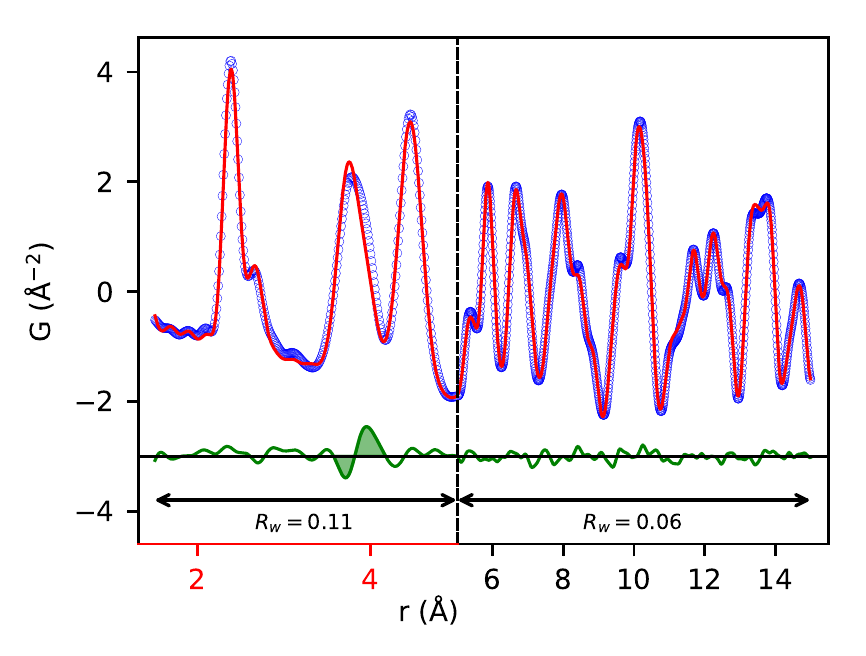}
\caption{\label{fig:poortetragonalfit} The signature of local nematicity. PDF of best-fit model (red solid line) to observed data (open circles) from \fese\ at 300~K, fitted up to $r = 50$~\AA\ using the tetragonal model (overall $R_w = 0.08$).
The fit is presented over the abbreviated range $1.5<r<15$~\AA\ for clarity.
The difference curve is shown displaced below (green solid line), and shaded to highlight the signature of local nematicity.
$R_w$ values over selected abbreviated $r$-ranges are presented to highlight low-$r$ misfit.
Different coloring of the x-axes of two panels signals that the scales are distinct to highlight the signal at low-$r$.
The y-axis scales are preserved between the two panels.
Importantly, there is also a subtle misfit of the Fe-Fe peak at 2.7~\AA\ (see text for details).}
\end{figure}

In Fig.~\ref{fig:poortetragonalfit} we show the PDF analysis results using the tetragonal model over the range $1.5<r<50$~\AA.
An abridged $r$-range is shown for clarity and the resulting structural parameters are summarized in Table~\ref{tab:strucparams}.
The fit is acceptable, with an overall $R_w = 0.08$, suggesting that the tetragonal model adequately describes the average structure of \fese\ at 300~K.
\bgroup
\def\arraystretch{1.5}%
\begin{table}[tbp]
\caption{\label{tab:strucparams} Structural parameters and fit details for the \fese\ sample at 300~K, obtained from PDF analysis using either the tetragonal or orthrhombic model.
Colors in pair distances are consistent with dashed lines in Fig.~\ref{fig:strucfig}}
\begin{tabular}{lcc}
{Parameter} & {Tetragonal} & {Orthorhombic} \\ \hline
$r$-range (\AA) & 1.5--50 & 1.5--5.0 \\
$R_w$ & 0.08 & 0.04 \\
$a$ (\AA) & 3.771 & 5.32 \\
$b$ (\AA) & -- & 5.41 \\
$c$ (\AA) & 5.520 & 5.50 \\
$z$ & 0.266 & 0.235 \\
$r$(Fe--Se) (\AA) & 2.391 & 2.391 \\
$r$(Fe--Fe) (\AA) & \textcolor{linebrown}{2.667} & \textcolor{linebrown}{2.661} \\
 & -- & \textcolor{linebrown}{2.710} \\
$r$(Se--Se) (\AA) & \textcolor{lineblue}{3.710} & \textcolor{lineblue}{3.709} \\
 & -- & \textcolor{lineblue}{3.742} \\
 & \textcolor{linegreen}{3.771} & \textcolor{linegreen}{3.795} \\
 & \textcolor{linered}{3.970} & \textcolor{linered}{3.944} \\
 & -- & \textcolor{linered}{3.974} \\ \hline
\end{tabular}
\end{table}
\egroup
However, a feature at $\sim$3.9~\AA\ in the local structure portion of the PDF is not well reproduced by the tetragonal model.
The misfit on this peak is associated with a larger $R_w$ of 0.11 over the range $1.5<r<5$~\AA, compared to $R_w = 0.06$ over the range $5<r<15$~\AA, as highlighted in Fig.~\ref{fig:poortetragonalfit}.

The PDF misfit in a single low$-r$ peak (Fig.~\ref{fig:poortetragonalfit}) is reminiscent of the fingerprint of a SRO ODL state in \cis~\cite{bozin_local_2019}.
In that system, a large misfit was observed in the first Ir-Ir PDF peak at $r=3.5$~\AA\ when the PDF was fit with the reported cubic average structure, with an absence of significant misfit at other places in the PDF.
This discrepancy was explained by the presence of fluctuating local symmetry-broken tetragonal domains present in the average cubic phase, that act as a precursor for the long-range ordered symmetry broken ground-state seen at low temperature in the \cis\ system.

In our \fese\ system the corresponding reminiscent misfit feature occurs at $r=3.9$~\AA\, corresponding to a composite peak originating primarily from Se-Se nearest-neighbor (NN) correlations.
The \fese\ tetragonal model allows for three distinct Se-Se NN distances.
The first set of NN Se-Se interatomic vectors spans the sheet created by the FeSe$_4$-tetrahedra between Se pairs with distinct $z$ coordinates (red dashed lines in Fig.~\ref{fig:strucfig}).
The second spans the space between different sheets, again involving Se pairs with distinct $z$ coordinates (blue dashed lines in Fig.~\ref{fig:strucfig}).
The third connects Se pairs with identical $z$ coordinates, either fully above or below the Fe species in the same layer (green dashed lines in Fig.~\ref{fig:strucfig}).

These symmetry constraints are lifted for some but not all of these interatomic vectors when the orthorhombic distortion occurs.
The NN Se-Se interatomic vectors that connect Se ions with distinct $z$ coordinates, represented by red and blue dashed lines in Fig.~\ref{fig:strucfig}, all lie along the orthorhombic unit cell axes (i.e., the basal-plane projection of the interatomic vectors are parallel to $a$ and $b$, respectively).
The orthorhombic distortion allows these pair distances to become distinct, with the magnitude of their difference dictated by the magnitude of the orthorhombic distortion.
On the other hand, the NN Se-Se interatomic vectors with zero $z$ component, represented by green dashed lines in Fig.~\ref{fig:strucfig}, are parallel to the diagonals of the orthorhombic $a$-$b$ plane and they remain identical following the orthorhombic distortion.
The reduction of symmetry associated with the orthorhombic distortion therefore results in an increase from three to 5~distinct Se-Se NN distances.
\ins{The feature at $r=3.9$~\AA\ also contains a Fe-Fe next nearest neighbor (NNN) contribution, but this peak is smaller than the Se-Se contributions by a factor of about five, and the orthorhombic distortion does not split the two Fe-Fe NNN distances, as they span the unit cell diagonals.}

The feature at $\sim$3.9~\AA\ in the difference curve in Fig.~\ref{fig:poortetragonalfit} suggests intensity is shifted to the right in the measured PDF compared to the model.
This indicates that the tetragonal model PDF under-represents higher-$r$ Se-Se NN pair correlations, while lower-$r$ Se pair correlations are over-represented.

Drawing from the previously mentioned \cis\ case~\cite{bozin_local_2019}, we seek an explanation in terms of a SRO locally orbital degeneracy lifted (SRO-ODL) state which persists to temperatures well above the observed long-range OO temperature $T_o$.
In \cis, the local structure of the material in the SRO-ODL state was well explained by the symmetry broken model, while the high-$r$ region was consistent with the high-temperature non-symmetry-broken structural model.
We therefore followed a similar analysis strategy as adopted in the \cis\ case.

\begin{figure}[tpb]
\includegraphics[width=3.5in]{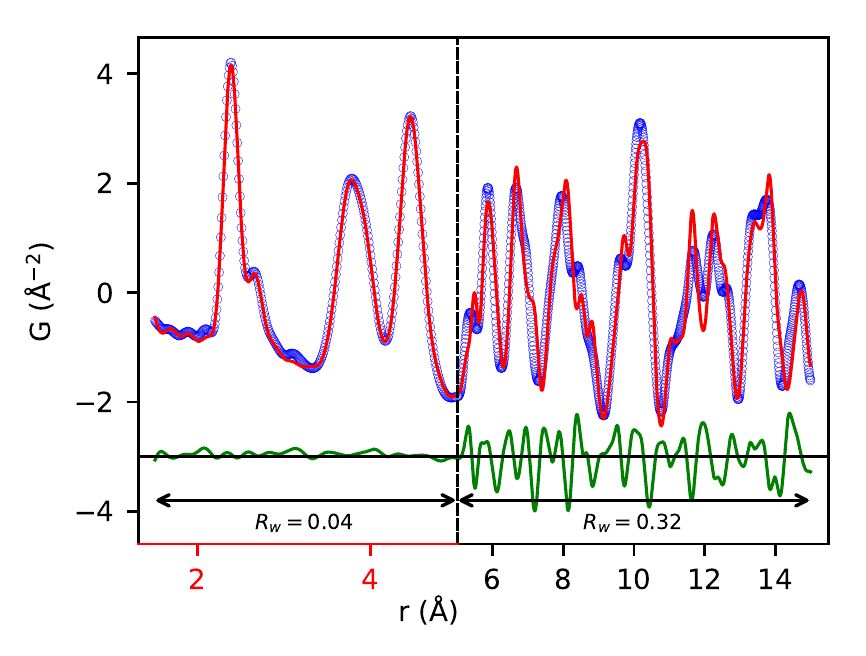}
\caption{\label{fig:orthorhombicfit} PDF of best-fit model (red solid line) and observed data (open circles) from \fese\ at 300~K, fitted up to $r= 5$~\AA\ using orthorhombic (nematic) symmetry constraints ($R_w = 0.04$).
The difference curve is shown displaced below (green solid line).
$R_w$ values over selected $r$-ranges are presented to highlight the discrepancy between the optimal local structure and the observed average structure.
Different coloring of the x-axes of two panels signals that the scales are distinct to highlight the signal at low-$r$.
The y-axis scales are preserved between the two panels.
\ins{The orthorhombic model provides observably better account of the Fe-Fe NN PDF peak as compared to the tetragonal model (Fig.~\ref{fig:poortetragonalfit})}}
\end{figure}

As mentioned, long-range OO in \feses\ manifests structurally as a tetragonal to orthorhombic transformation, which involves an increase in the number of distinct Se-Se NN distances.
Our analysis applying the low temperature orthorhombic model to the low-$r$ ($1.5<r<5$~\AA) region of the 300~K PDF is shown in Fig.~\ref{fig:orthorhombicfit}, with resulting structural parameters presented in Table~\ref{tab:strucparams}.
This model successfully removes the reminiscent misfit feature that is observed in this $r$-region with the tetragonal model, yielding an $R_w$ nearly $3\times$ lower than the tetragonal model, ($R_w = 0.04$ vs. 0.11) over the range $1.5<r<5$~\AA.
This strongly suggests that at 300~K the local structure is orthorhombic and is a SRO-ODL state.

Compared to the average structure, the \rout{local} \ins{orthorhombic} distortion does not alter the Fe-Se distance, but does split \rout{Fe-Fe and} the NN Se-Se \ins{and NN Fe-Fe} pair distances significantly, while also shifting them to higher-$r$.
The misfit in Fig.~\ref{fig:poortetragonalfit} \rout{shows up primarily in}\ins{is more apparent in} the \ins{NN} Se-Se PDF peak at $\sim$3.9~\AA\, rather than the \ins{NN} Fe-Fe PDF peak at $\sim$2.7~\AA\ due to the stronger scattering power of Se and a larger Se-Se multiplicity, resulting in a peak that is roughly four-times stronger than the \ins{NN} Fe-Fe peak.

A similar sized distortion \ins{in the Fe and Se sub-lattices} thus results in a much larger signature in the difference curve \ins{associated with Se-Se features}.
However, and importantly, \rout{close}\ins{careful} inspection of the Fe-Fe peak at 2.7~\AA, evident as a well resolved shoulder to the Fe-Se peak in Fig.~\ref{fig:poortetragonalfit}, indicates that this peak is \ins{observably} broader in the data than the model.
This misfit, although subtle, is consistent with the presence of an ODL state in the Fe-Fe sublattice.
\ins{While such a misfit alone could easily escape attention, the broader context of the large residual at  $\sim$3.9~\AA\ highlights this region.}

As in the \cis\ case, if we calculate the PDF of the orthorhombic model, which reproduces the low-$r$ region well, over a much wider $r$-range (up to 15~\AA), it results in a poor agreement to the measured PDF compared to the non-symmetry broken tetragonal model (Fig.~\ref{fig:orthorhombicfit}, $R_w = 0.32$ for $5<r<15$~\AA\ compared to $R_w = 0.06$ for the tetragonal model over this range).
This implies that the distortion has orthorhombic short-range character.

\begin{figure}[tpb]
\includegraphics[width=3.5in]{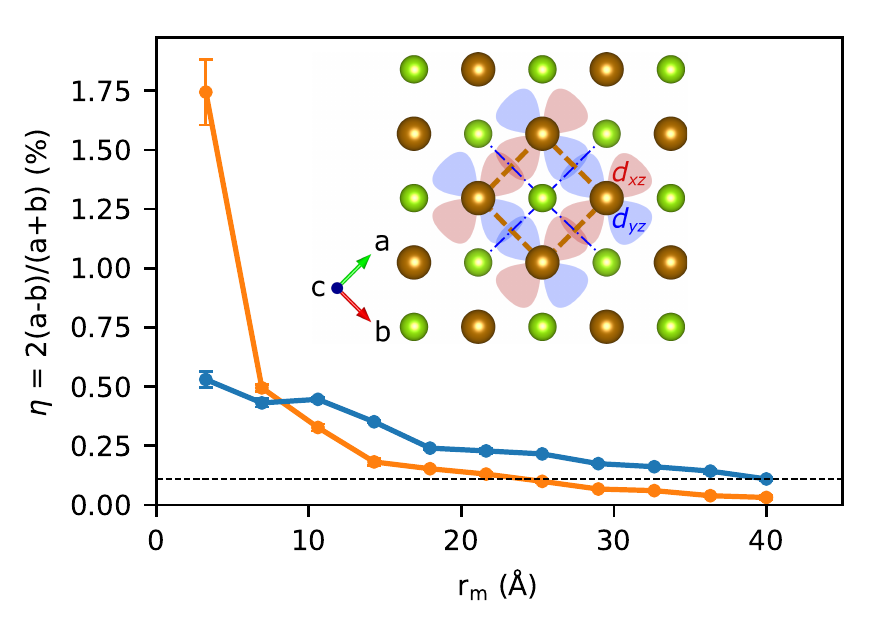}
\caption{\label{fig:correlationlength} The measured orthorhombic distortion obtained by using an orthorhombic model over a narrow $r$-range, $\eta$, is plotted vs the PDF fit window center, $r_m$, for data measured at 300~K (orange line) and 84~K (blue line).
The value of $\eta$ at 84~K as extracted from the average structure is represented by the dashed horizontal line.
Inset is a top view of the \fese\ crystal structure, schematically showing the projected orientations of the Fe 3$d_{yz}$ and 3$d_{xz}$ orbitals implicated in this work and elsewhere.
Atom and dashed line colors are consistent with Fig.~\ref{fig:strucfig} and the axes labels are consistent with the orthorhombic setting.
Full PDF fits associated with the distortion magnitudes in this plot are shown in Fig.~\ref{fig:boxcarPDFS} in the SI}
\end{figure}

Given that we have found local orthorhombic domains in a globally tetragonal crystal, we would like to extract the size, or structural coherence, of the domains.
To do this we use box-car style PDF fits~\cite{qiu;prl05,bozin_cuir1_2014-1} where \rout{small} \ins{the fits are carried out over a number of variable} $r$-range \ins{regions individually.} \rout{(the box-cars) are fit, with the centroid of the box-car varying in $r$.}  
\ins{The protocol represents an extension of the fitting shown in Fig.~\ref{fig:orthorhombicfit}, wherein the ``box-car'' consisted of the $r$-range of 1.5-5~\AA, and a centroid $r_m = 1.75$~\AA.
Each distinct $r$-range fit then provides structural information relevant to a distinct distance window, allowing for a quantification of the variation of local structure as a function of $r$ (the correlation length of local features).}
The \rout{results}\ins{orthorhombic distortion $\eta$ as a function of the box-car centroid, $r_m$,} is shown in Fig.~\ref{fig:correlationlength}, \ins{demonstrating the robustness of the approach (see SI and Fig.~\ref{fig:boxcarPDFS} for details)}

Surprisingly, our analysis reveals that the local distortion above $T_s$ is about 1.75~\%, more than $3\times$ larger than any observed distortion in the average structure below $T_s$.
The magnitude of the measured distortion drops rapidly with increasing $r$ (Fig.~\ref{fig:correlationlength}) until, at length scales greater than 40~\AA, it is effectively zero and the average tetragonal structure is recovered.

Interestingly, even below $T_s$ (at $T=84$~K) the local orthorhombicity is larger than that observed in the average structure at this temperature, represented by the dashed horizontal line in Fig.~\ref{fig:correlationlength}.
At this temperature, the local distortion also decreases with increasing $r$, converging to the measured average orthorhombic structure at length scales greater than 40~\AA.
Based on the proposed SRO-ODL explanation, the persistence of a larger-than-average local orthorhombicity below $T_s$ suggests that the long-range ordering of the ODL state is not perfect.



The xPDF results presented here build on and extend previous studies~\cite{iye_emergence_2015, frandsen_widespread_2018} showing similarly local $C_4$ symmetry breaking in the 122 family of iron pnictide superconductors, suggesting that such fluctuations may be widespread.

\rout{Further, our structural findings corroborate spectroscopic studies which have observed electronic non-degeneracy in \fese\ well above $T_s$.}
\ins{Further, our findings provide a \instwo{direct structural} rationale for \instwo{anomalous line width and splitting observed in NMR spectra}~\cite{wiecki_nmr_2017} \instwo{as well as unexpected energy gaps observed in the electronic spectra}}~\cite{wen_gap_2012, zhang_observation_2015, wang_robust_2017} \ins{pointing towards electronic non-degeneracy in \fese\ well above $T_s$.}
The 3$d_{yz}$ and 3$d_{xz}$ Fe orbitals implicated in this local degeneracy lifting are oriented along the distorted orthorhombic direction (see e.g. Fig.~\ref{fig:correlationlength} inset) and participate in orbital-selective Cooper pairing~\cite{sprau_discovery_2017, kostin_imaging_2018} enhanced by nematic order~\cite{yu_orbital_2018}.
Thus the xPDF discovery of a local ODL state deep in the high temperature tetragonal regime not only defines a local structural fingerprint of the fluctuating nematic state, but also serves to reinforce the role of the orbital sector and its coupling to structure within the physics of iron-based superconductors.
\ins{Local orbital overlaps governed  by nematic fluctuations appear to be essential for slow dynamic correlations in FeSe and related materials}~\cite{Zaliznyak10316, konstantinova2019photoinduced}.
\ins{Our observation of a broken symmetry state at high temperature suggests that the global symmetry breaking structural transition at T$_s$ occurs through crystallization of the preformed liquid-like orbital state, a description which challenges present understanding of this system and invites further theoretical considerations.}

To that effect, it will be of particular interest to explore the evolution of this local ODL state away from the \fese\ parent, across the phase diagrams of e.g. FeSe$_{1-x}$S$_{x}$ and FeSe$_{1-x}$Te$_{x}$ and in the regime where long range symmetry breaking transition is not observed at any temperature.
Such studies would further inform the debate of the importance of nematic fluctuations for the superconductivity itself.



%

%
%
\begin{acknowledgments}
This work was supported by U.S. Department of Energy, Office of Science, Office of Basic Energy Sciences (DOE-BES) under contract No. DE-SC0012704.
X-ray PDF measurements were conducted on beamline 28-ID-1 of the National Synchrotron Light Source II, a U.S. Department of Energy (DOE) Office of Science User Facility operated for the DOE Office of Science by Brookhaven National Laboratory under Contract No. DE-SC0012704.
\end{acknowledgments}

\bibliography{billinge-group,abb-billinge-group,everyone,18sb_nematicityFeSe}
\bibliographystyle{apsrev}
\vfill\newpage
\renewcommand\thefigure{S\arabic{figure}}
\setcounter{figure}{0}
\section{Supplementary Information}
\subsection{Electrical transport properties}
\label{sec:rho}
\begin{figure}[tpb]
\includegraphics[width=3.5in]{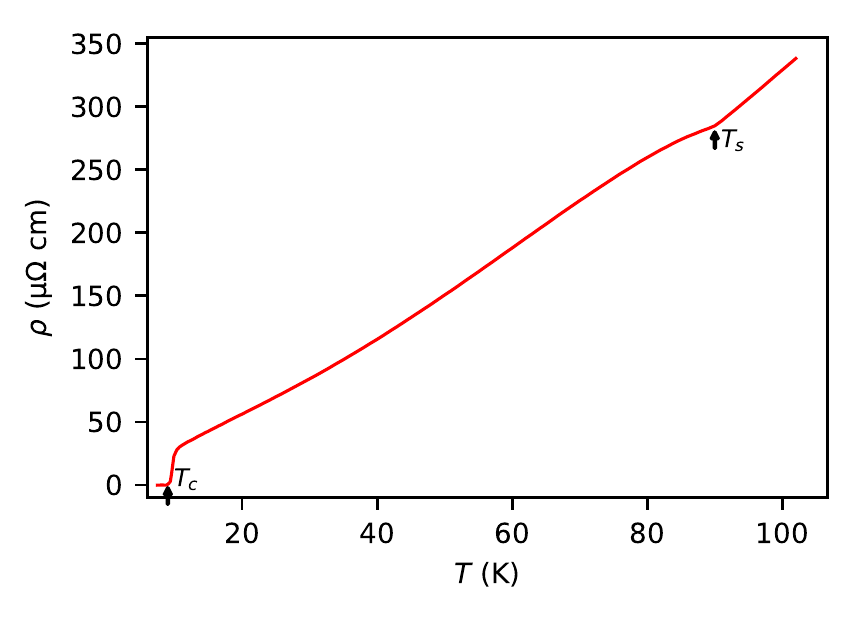}
\caption{\label{fig:transport} Temperature dependence of resistivity, $\rho$, of \fese\ sample.
Relevant temperatures $T_s$ and $T_c$ are labeled and marked with arrows.}
\end{figure}

Temperature dependent electrical resistivity measurement were conducted in a Quantum Design PPMS-9.
Electrical contacts were made using standard four-probe configuration.
Results are shown in Fig.~\ref{fig:transport}, where $T_c$ and $T_s$, at 10~K and 90~K, respectively, are labeled.
\subsection{xPDF fitting details}
\label{sec:PDF}
The xPDF data were fit with the \pdfgui~\cite{farro;jpcm07} program using the orthrhombic or tetragonal \fese\ models as described in the text.
In both cases, the lattice parameters were refined, along with the single $z$ atomic position parameter, a correlated motion parameter $\delta_1$~\cite{proff;jac99}, thermal parameters, and a scale parameter.
Anisotropy in the thermal parameters was allowed based on the symmetry of the model, with the exception that off-diagonal terms were fixed at zero.
To quantify the impact on the PDF of $Q$-space instrument resolution, $Q_{damp}$ and $Q_{broad}$~\cite{toby;aca92} were refined using PDF data collected from a nickel standard.
The refined values of $Q_{damp} = 0.042$~\AA$^{-1}$ and $Q_{broad} = 0.010$~\AA$^{-1}$ are fixed in all subsequent refinements to the xPDF data.

\begin{figure}[tpb]
\includegraphics[width=3.5in]{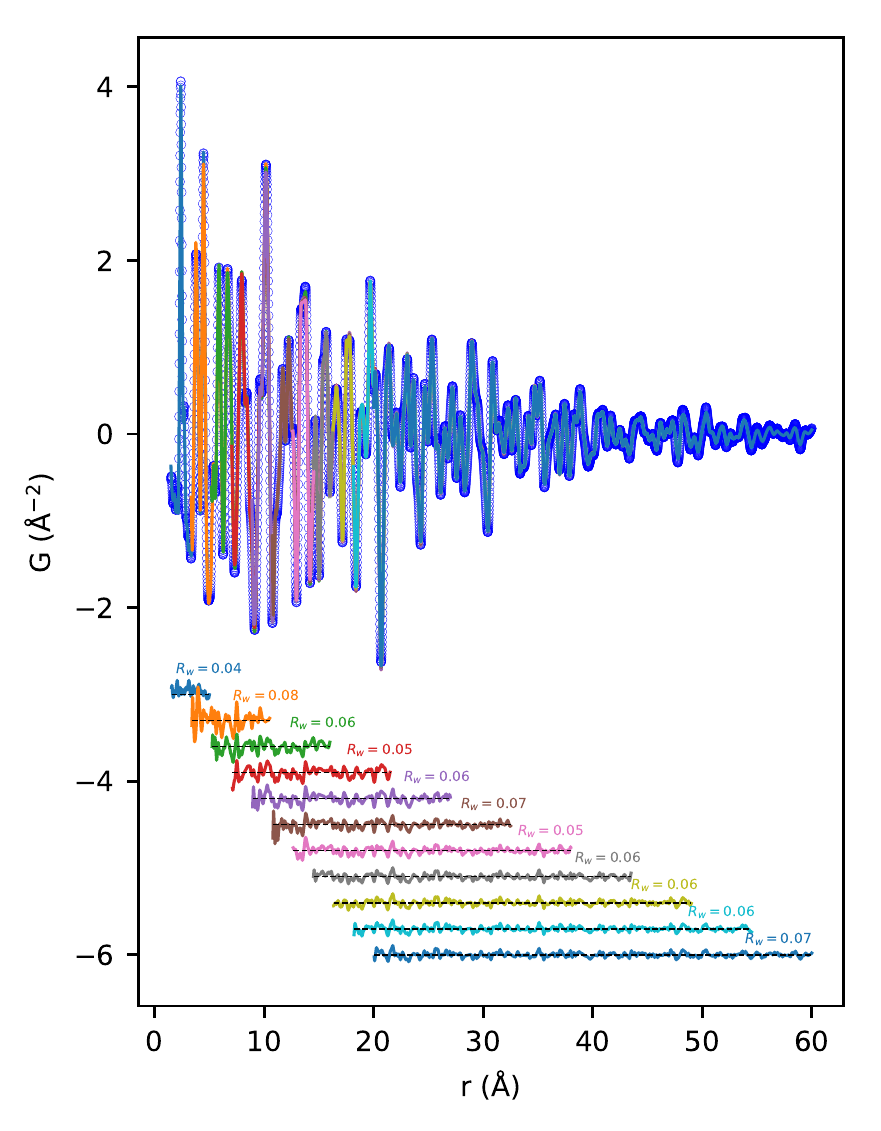}
\caption{\label{fig:boxcarPDFS} Sliding box-car style PDF fits, with the models (overlapping colored solid lines) and observed data (open circles) from \fese\ at 300~K, fitted over different (shown) $r$-ranges using orthorhombic (nematic) symmetry constraints. Difference curves for each $r$-range are offset vertically, and match the color of their model curve.
$R_w$ for each $r$-range is listed in the corresponding color.
 These fits were used to quantify the correlation length of the local orthorhombic distortion as shown in Fig.~\ref{fig:correlationlength}}
\end{figure}

\ins{Box-car type fitting was done with 11 different fitting windows with $r_{min}$ varying from $1.5-30$~\AA\ in steps of 2.85~\AA\ and $r_{max}$ varying from $5-70$~\AA\ in steps of 6.5~\AA. PDF fitting results are shown in Fig.~\ref{fig:boxcarPDFS}. 
We have chosen these box sizes such that we explore correlation ranges spanning from one unit cell up to the field of view available in our data
The local orthorhombic distortion is not found only in an $r$-range of 1.5-5~\AA, but also in 10 additional PDF fits spanning multiple unit cells, albeit with a diminished amplitude, reflecting the limited limited correlation length of this effect.}

\end{document}